# Thick GEM-like hole multipliers: properties and possible applications

R. Chechik[a], A. Breskin, C. Shalem and D. Mörmann

*Department of Particle Physics, The Weizmann Institute of Science, 76100 Rehovot, Israel*

**Abstract**

The properties of thick GEM-like (TGEM) gaseous electron multipliers, operated at 1-740 Torr are presented. They are made of a G-10 plate, perforated with millimeter-scale diameter holes. In single-multiplier elements, effective gains of about $10^4$, $10^6$, and $10^5$ were reached at respective pressures of 1, 10 Torr isobutane and 740 Torr Ar/5%$CH_4$, with pulse rise-times in the few nanosecond scale. The high effective gain at atmospheric pressure was measured with a TGEM coated with a CsI photocathode. The detector was operated in single and cascaded modes. Potential applications in ion and photon detection are discussed.

*Keywords:* GEM; gaseous electron multipliers; low-pressure avalanche multiplication; photon detectors; ion detectors.

## 1. Introduction

Gas avalanche multiplication within small holes is attractive because the avalanche-confinement in the hole strongly reduces secondary effects, and they provide true pixilated radiation localization. Hole-based multiplication has been the subject of numerous studies in a large variety of applications. Recently, proportional gas multiplication was demonstrated in glass channel plates (CP) [1, 2]. The most attractive and extensively studied hole-multiplier is the Gas Electron Multiplier (GEM) [3], made of 50-70-µm diameter holes etched in a 50-µm thick metalized Kapton foil. It operates in a large variety of gases including noble-gas mixtures, providing a gain of $\sim 10^4$ in a single element and gains exceeding $10^6$ in a cascade of 3-4 elements [4, 5]. The avalanche process is fast (typical rise-time of a few ns) and generally free of photon-mediated secondary effects [6]. In addition to its use for particle tracking and in TPCs, the GEM is also efficiently coupled to gaseous or solid radiation converters, resulting in a large variety of GEM-based radiation detectors developed for imaging of x-rays, neutrons and UV-to visible light [7].

The success of the GEMs and glass CPs triggered the concept of a coarser structure made by drilling millimetric holes in a 2mm thick Cu-plated G-10 printed circuit board (PCB) [2, 8]. These yielded gains of $10^4$ in Ar/5% isobutane and in pure Xe and $10^3$ in pure Xe, when combined with a CsI photocathode (PC).

The thick GEM-like (TGEM) multiplier investigated here is a simple and robust detector, with millimetric pixilation. In combination with appropriate radiation converters, it has potential applications for the detection of light, neutrons, x-rays, charged particles etc. We describe its operation properties at very low gas pressures, in view of low-energy ion detection, and at atmospheric pressure (with a CsI photocathode), for Cherenkov UV-photon imaging.

## 2. Experimental setup and procedures

The TGEMs were manufactured with standard PCB technology of precise drilling and Cu etching, out of double-face Cu-clad G-10 plate, of 1.6 – 3.2 mm thick. A gap of 0.1mm was kept between the rim of the drilled hole and the edge of the etched Cu pattern (fig. 1).

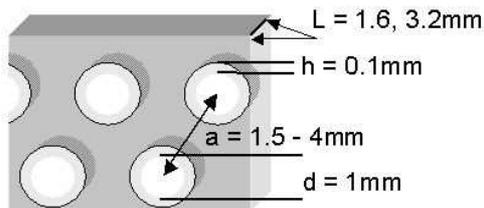

Figure 1. A scheme of the thick GEM-like multiplier.

The effective gain was measured in a "current" mode with a continuous Hg(Ar) UV-lamp irradiating either a semitransparent (ST) CsI PC, placed at 8mm above the TGEM, or a reflective (REF) PC deposited on its top face. In the first, normalization, step the photo-induced current was recorded on the TGEM with its two sides interconnected (ST PC, fig. 2a), or on the mesh installed 6.5mm above the TGEM (REF PC, fig. 2b), with a drift field $E_{drift}$ resulting in photoelectron collection plateau. In the second step, with the ST PC we maintained the same $E_{drift}$ field in the gap above the TGEM and recorded the current at the bottom electrode as function of the voltage difference between its both sides, keeping a slightly reversed field in the gap below the TGEM (fig. 2c). With the REF PC we maintained $E_{drift}=0$ for optimal electron extraction and focusing [9] and recorded the current in the same way (fig. 2d). By dividing the currents from the second measurement to that of the first one, we obtained the absolute effective gain curve; it represents the product of the true gain in the holes and the efficiency to focus the photoelectrons into the holes. The latter also varies as function of the field in the holes.

Some effective-gain measurements were carried out in a "pulse" mode, with a pulsed $H_2$ lamp providing multi-photon bursts. Here, the charge-signal's pulse-height was recorded at the bottom electrode, as a function of the voltage across the TGEM; the field configurations were kept as explained above. The results of the "pulse" mode measurements were normalized to that of the "current" ones. Fast current pulses of single- and multi-photons were also recorded from the bottom electrode with a fast (1ns rise-time) amplifier.

In a cascaded-mode operation, two TGEMs were mounted at a distance of 10mm (fig. 2e); currents were recorded at the bottom electrode of the second TGEM, maintaining a transfer field between the elements and a slightly reversed field in the gap below. This effective gain in cascaded mode, involves the product of the true gains in both TGEMs, the efficiency to focus electrons into both and the efficiency to extract electrons from the first one into the gap between them. These efficiencies are a-priory unknown and must still be optimized; they depend on electron diffusion and on the different electric fields configurations [10].

## 3. Electric field calculations

MAXWELL software package [11] was used to study the role of the TGEM geometrical parameters and its expected performance in terms of electron transport into the holes. We can summarize the following: the field in the hole's center does not depend on the distance between adjacent holes but rather on the TGEM thickness. The field at the center of a 1.6mm thick TGEM, operated at atmospheric pressure under a maximal voltage difference of 3 kV, is of 17 kV/cm; it is ~6 times smaller than the corresponding field in a standard GEM.

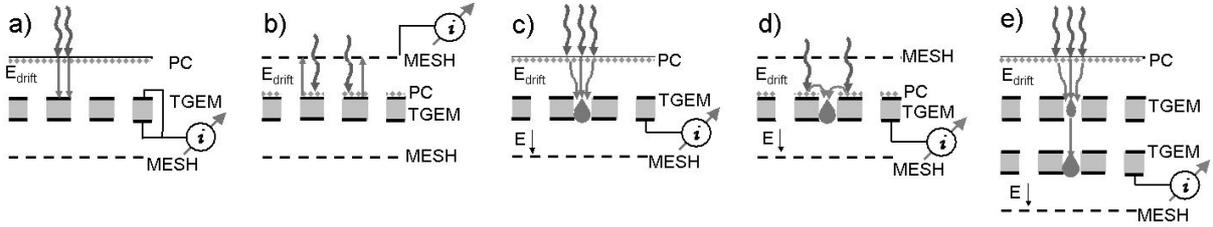

Figure 2. Schemes for normalization and absolute effective-gain measurement in semitransparent (a,c), reflective (b,d) and double TGEM (a,e) modes.

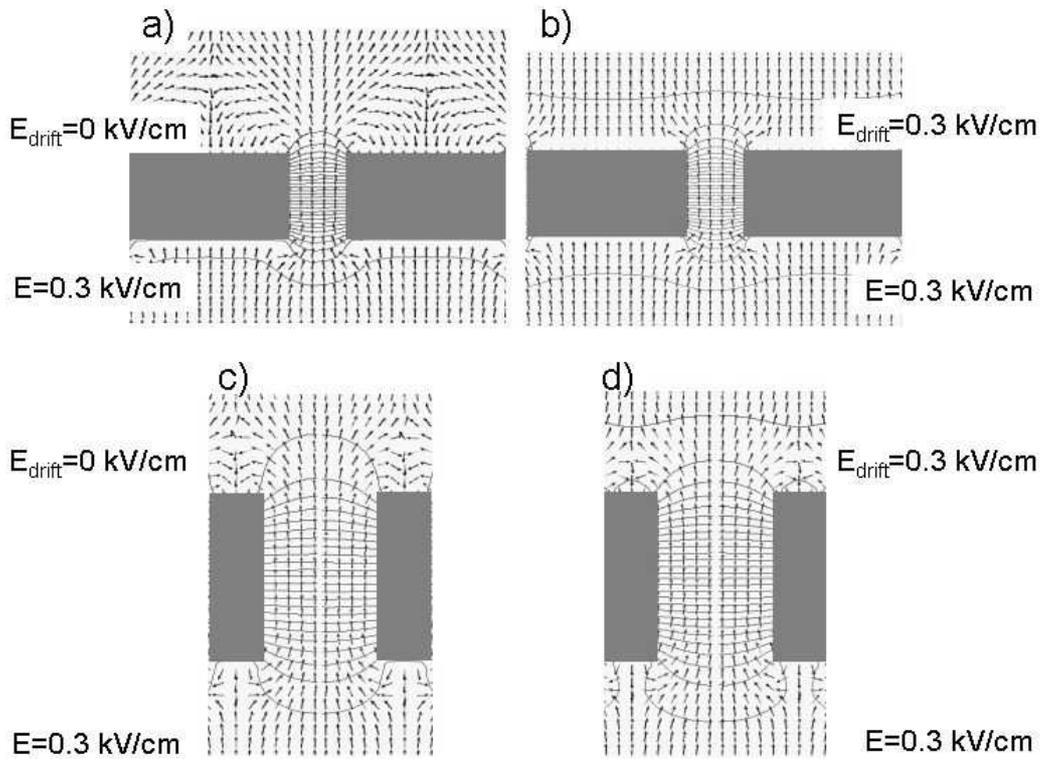

Figure 3. Equipotential- and electric-field lines calculated for TGEMs of 1.6mm thick and 1mm holes. Results are shown for pitch=4mm (a, b) and 1.5mm (c, d) and for $E_{drift}$=0 and 0.3kV/cm. Notice the different scales in the schemes.

The field penetration outside the holes is also very small, of about 0.3kV/cm at atmospheric pressure. This implies that $E_{drift}$ in the gap above the TGEM has to be very carefully optimized for maximal focusing of electrons into the holes. A too large field will transport the electrons into the metallic surface between the holes, while a too low field implies large electron diffusion, and thus inefficient focusing. Fig. 3 shows a vectorial field-map representation, depicting the electric-field direction and strength. It is shown for 1 mm holes in 1.6 mm thick TGEM having 1.5 and 4 mm pitch, under two configurations: $E_{drift}$=0 and $E_{drift}$ = 0.3 kV/cm. For a 4 mm pitch and $E_{drift}$ = 0.3 kV/cm (fig. 3b) there is no efficient focusing into the holes. With $E_{drift}$=0 (fig. 3a), the filed-lines direction permits focusing of electrons originating either from the gap above the TGEM or from a converter deposited on its top electrode, into the holes. However, since the field strength on the TGEM surface is practically 0, while at atmospheric-pressure operation a field ≥1 kV/cm is required to overcome backscattering losses [12], it will not provide efficient electron extraction. For the 1.5 mm pitch, the field-line configuration is more adequate; with $E_{drift}=0$ the field strength at the top-TGEM electrode is >1kV/cm at atmospheric pressure operation conditions, which guarantees minimal backscattering losses.

## 4. Results

All results presented in this chapter refer to a 1.6 mm thick TGEM, with 1mm holes and 1.5mm pitch.

Fig. 4a depicts the effective gain measured at 740 Torr Ar/5% $CH_4$, of a single TGEM with ST and REF CsI PCs. An effective gain ~$10^5$ was measured with both. For the same voltage difference across the TGEM the ST mode effective gain is by a factor 10 smaller than the REF one; it indicates upon a ~10-fold poorer electron focusing into the holes. Moreover, in the double-TGEM mode the same effective gain was reached as with a single-TGEM under the same voltages across each TGEM, indicating upon considerable electron losses in the transfer between them. These observations point at a low efficiency of electron focusing from the gap into the holes, with $E_{drift}$ in the range of 0.3-0.5kV/cm in this gas mixture.

Fig. 4b depicts the effective gain measured at 1 Torr of isobutane. The total effective gain is $10^4$ (at $\Delta V_{TGEM}$>450V the gain curve deviates from proportionality) and the results from the ST and the REF PC modes, measured independently, seem to overlap particularly in the higher gain range; this indicates at an identical electron focusing efficiency, which is most probably high. In the double-TGEM operation mode the total effective gain is also $10^4$, but obtained under a much smaller voltage difference across each element; it indicates at efficient electron transfer between the two multipliers. Both observations point at an efficient focusing from the gap into the holes, under $E_{drfit}$ =20-25V/cm in this gas and pressure.

Fig. 4c depicts the data measured at 10 Torr of isobutane; the results are very similar to those at 1 Torr, with a total effective gain approaching $10^6$, though there is a slight difference between the ST and REF PC modes. Similar observations and conclusions hold for the double-TGEM operation at 10 torr, and for the electron focusing efficiency under $E_{drift}$ of ~200V/cm in this gas and pressure.

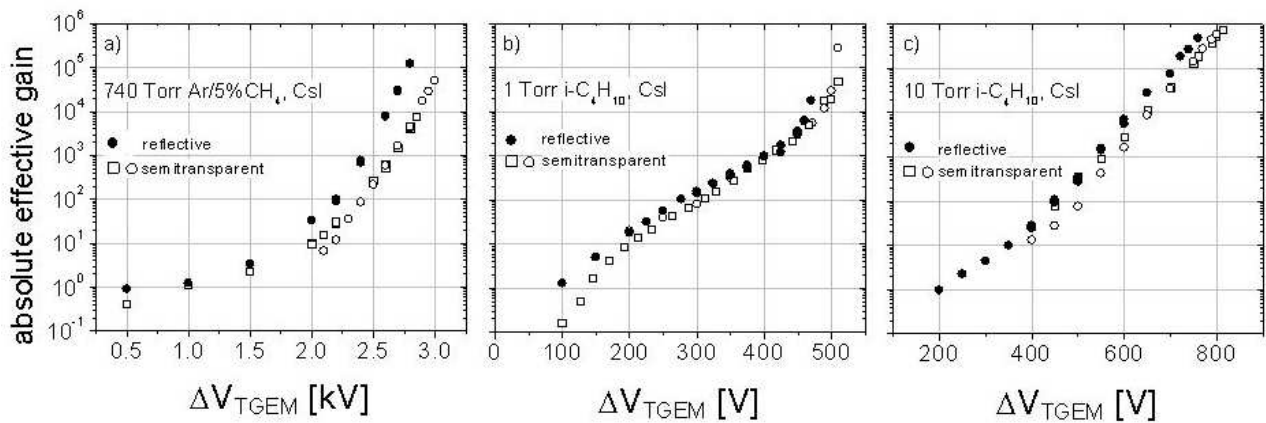

Figure 4. Absolute effective gain measured at 740 Torr Ar/5%$CH_4$ (a) and at 1 (b) and 10 (c) Torr i-$C_4H_{10}$, in reflective mode (closed circles) and in semitransparent "current" (open squares) and "pulse" (open circles) modes.

Fast pulses recorded in a single-TGEM with single- and multi-photon bursts show respective rise-time of 10, 5 and 3.5 ns, at 740, 10 and 1 Torr (Fig. 5).

## 5. Summary and applications

Thick GEM-like elements, manufactured in standard PCB technology are robust multipliers; they have astonishingly high gains: $\sim 10^5$ at atmospheric pressure and $\sim 10^4$-$10^6$ at 1-10 Torr. The signals are fast, with rise times in the ns range. The operation of cascaded TGEMs was demonstrated at low pressures, where the electron focusing efficiency into the holes is probably high. This operation mode is important for the suppression of secondary effects, and in particular the reduction of ion back-flow [13]. We have stressed the importance of the electron transport issue; it requires further studies for the optimization of the geometry and the electric fields.

TGEMs operated at atmospheric pressure have many potential applications. An example is in Ring Imaging Cherenkov (RICH) detectors, where photon detection over several square meters is required, generally with moderate (millimetric) localization accuracies; pad-readout with modern VLSI electronics [14, 15] permits operation at gains in the $10^4$-$10^5$ ranges. The TGEM with a REF PC is an attractive robust solution, with very low sensitivity to background ionizing radiation [9]. Other applications could be the detection of scintillation light in large noble-gas detectors, e.g. in search for WIMPS [8] and in x-ray and neutron detectors with appropriate converters, where the gain reached in a single TGEM could be sufficient. For such applications detector optimization is in course.

The operation at very low pressures, of 1 Torr or less, is interesting for ion detection; an example is tracking nanodosimetry, where radiation-induced ions in a very dilute gas are detected and localized to provide the ionization track structure. The dilute gas provides a factor $10^6$-dimension expansion of the radiation-track image, as compared to condensed matter such as tissue, resulting in track-structure data relevant to the scale of the DNA molecule [16]. The TGEM coupled to an appropriate ion converter could be the basis for such an ion-imaging detector. Low-pressure TGEMs could also become adequate economic imaging and timing detectors for heavy ions in nuclear- and atomic-physics applications.

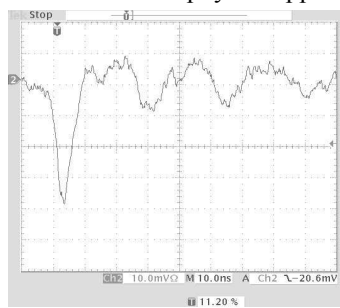

Figure 5. A fast single-photon pulse, of 5ns rise-time, at 10 Torr i-$C_4H_{10}$, gain~$6\times 10^5$. The after-pulses are due to electronic noise.

## Acknowledgments

The work was supported in part by the Israel Science Foundation, project No. 151-01 and by the Binational Science Foundation, project No. 2002240. C.S. was supported in part by the Fund for Victims of Terror of the Jewish Agency for Israel. A.B. is the W.P. Reuther Professor of Research in the peaceful use of Atomic Energy.